\documentstyle[aps,prb,epsf,multicol]{revtex} 
\input epsf
\newcommand{\dir}{Figs}

\newcommand{\fig}[4]
{
     \noindent
     \unitlength=1mm
     \begin{picture}(#2,#3)
     \put(10,0){\leavevmode \epsfxsize=#2mm \epsffile{\dir/#1}}
     \end{picture}
   \noindent
#4
}
\newcommand{\rr}{ {\bf r} }
\newcommand{\ru}{ {\bf \hat{r}} }
\newcommand{\rg}{ {\hat{\gamma}} }
\newcommand{\nn}{ {\bf n} }
\newcommand{\uu}{ {\bf u} }
\newcommand{\QQ}{ {\bf Q} }
\newcommand{\II}{ {\bf I} }
\newcommand{\RR}{ {\bf R} }

\newcommand{\bt}{ {\overline{\theta}} }
\begin{document} 

\newcommand{\CCsnap}
{
\caption{
Configuration snapshots (left half of the simulation box) at grafting 
densities $\Sigma = 0.34/\sigma_0^2$ and (top) and 
$\Sigma = 0.84/\sigma_0^2$ (bottom). Solvent particles are dark,
and chain particles are bright. 
}
\label{fig:snapshots}
\bigskip
}

\newcommand{\CCprofiles}
{
\caption{ 
Profiles of the order parameter $S$ (dotted line) and the $z$ component 
(dashed line) of the director $\nn$ as a function of the distance from
the surface $z$ for grafting densities $\Sigma = 0.34/\sigma_0^2$ 
(a) and $\Sigma = 0.84/\sigma_0^2$ (b).  
Also shown for comparison is the density profile of chain particles 
(thin dashed line) and the density of chain end particles (thin solid
line) in units of $1/\sigma_0^3$.
}
\label{fig:profiles}
\bigskip
}

\newcommand{\CCcartoon}
{
\caption{
Surface region of the chain layer where the interactions between 
chains are reduced. See text for explanation.
}
\label{fig:cartoon1}
\bigskip
}

\newcommand{\CCboundary}
{
\caption{
Graphical solution of Eqn.~(\protect\ref{eq:bll_2}). The anchoring angle 
$\theta(L)$ is given by the point where $g(\theta)$ (dashed line) and 
$U(\theta)-U(\bt))$ (solid line) cross. If $g(0)$ is larger than $U(0)-U(\bt)$, 
the anchoring angle is zero. (The parameters leading to this particular plot were
$\zeta=0.1, v=1.7, \rg = 0.5, \delta K = 0$. Units are defined in the 
text~\protect\cite{footnote1}).
}
\label{fig:boundary}
\bigskip
}

\newcommand{\CCphdiag}
{
\caption{
Phase diagrams in $v-\zeta$ space (chain interactions vs. grafting density)
for $\rg/\sqrt{1+\delta K} = 0.5$ (a) and $\rg/\sqrt{1+\delta K} = 0.1$ (b).
Thin solid lines indicates tilting transition in the chain region, and thick
solid line the anchoring transition in the nematic fluid. Short dashed
line indicates the approximation (\ref{eq:low}) and long dashed line the
approximation (\protect\ref{eq:high}) (outside of the frame in the case of (b)).
Units are defined in the text~\protect\cite{footnote1}.
}
\label{fig:phdiag}
\bigskip
}

%

\title{An anchoring transition at surfaces with grafted 
       liquid-crystalline chain molecules}

\author{Harald Lange$^{\dag, \ddag}$ and Friederike Schmid$^{\dag}$}

\address{
$\dag$ Fakult\"at f\"ur Physik, Universit\"at Bielefeld, 
         33615 Bielefeld, Germany \\
$\ddag$ Institut f\"ur Physik, Universit\"at Mainz, 
55099 Mainz, Germany
}

\setcounter{page}{1}
\maketitle 
\tighten

\begin{abstract}

The anchoring of nematic liquid crystals on surfaces with grafted liquid
crystalline chain molecules is studied by computer simulations and within 
a mean field approach. The computer simulations show that a swollen layer
of collectively tilted chains may induce untilted homeotropic 
(perpendicular) alignment in the nematic fluid.
The results can be understood within a simple theoretical model.
The anchoring on a layer of mutually attractive chains is determined 
by the structure of the interface between the layer of chain molecules and 
the solvent. It is controlled by an interplay between the attractive chain 
interactions, the translational entropy of the solvent and its elasticity. 
A second order anchoring transition driven by the grafting density 
from tilted to homeotropic alignment is predicted.

\end{abstract}

\begin{center}
\end{center}

%
%

\section{Introduction} 
\label{sec:intro}

\begin{multicols}{2}
Nematic liquid crystals are fluids of elongated or oblate particles,
which lack translational order, but exhibit long-range orientational 
order\cite{degennes,chandrasekhar}: 
the molecular axes have a common preferred direction, the director. 
Surfaces and interfaces align the nearby molecules and thus favor
certain director orientations in the bulk~\cite{jerome}. 
This phenomenon, known as anchoring, plays a key role in the
design of liquid-crystal display devices~\cite{bahadur,schadt}. 
From a technological point of view, one is particularly interested 
in tailor-making surfaces which orient a liquid crystal with 
an arbitrary, well-defined anchoring angle.

In practice, alignment layers are often produced by mechanical rubbing of 
a polymer coated surface. The technique is simple and successful, yet it has 
drawbacks: Low tilt angles between the director and the surface normal
cannot be achieved easily, and the alignment mechanism is still not 
understood completely~\cite{toney,abbott}. As an alternative, Halperin and 
Williams  have suggested to use swollen brushes of liquid crystalline 
polymers as alignment layers\cite{avi1,avi2,avi3}. 
Their idea was to create a competition between the alignment favored by
the bare substrate and that enforced by the stretching of polymers in 
a dense brush. The interplay between the distortion
energy of the solvent director field and the conformational entropy of 
the brush was predicted to trigger a second order anchoring transition
between a phase with planar (homogeneous) alignment at low grafting 
densities and one with tilted alignment at higher grafting densities.
The tilt angle can be tuned by adjusting the grafting density.

Subsequently, Peng, Johannsmann and R\"uhe have undertaken first steps
towards an experimental realization of such a scenario\cite{peng1,peng2}.
A ``grafting-from'' technique~\cite{prucker} allowed to grow thick 
side-chain liquid crystalline polymer brushes from a substrate covered
with surface-attached initiators. A competition as required by Halperin and
Williams can be set up by covering the surface between the grafting 
sites with alkyl chains, which favor homeotropic alignment on the bare 
substrate. The resulting anchoring behavior has not yet been investigated. 
However, preliminary miscibility studies have shown that the liquid
crystalline brushes can only be swollen to a very limited extent by
low molecular weight nematic compounds, even if the latter are
chemically similar~\cite{peng3}. One would thus expect that the
scenario is altered: The brush and the nematic bulk are presumably
separated by an interface, which will contribute significantly to the 
anchoring properties of the alignment layer.

This aspect of surface anchoring on grafted liquid crystalline chains is 
explored in the present paper. We study brushes of chains which are too short 
or too stiff to develop hairpins and the like, and hence mainly follow
the director profile. If they attract each other, such short chains
exhibit collective tilt\cite{me1}. The contribution of the chain entropy to
the free energy is of minor importance. Instead, the elastic energy
of director distortions competes with a tendency of the chain ends at the 
surface of the chain layer to stand up, so that more solvent particles can
intrude into the interfacial region. As we shall see, the interplay 
between these two factors leads to a novel anchoring transition between 
a phase with tilted alignment and one where the layer of grafted chains
aligns the bulk vertically (homeotropic alignment), even though the chains 
inside of the layer are tilted. This is demonstrated with computer 
simulations and analyzed within a simple theoretical model.

Our paper is organized as follows: We discuss the simulation
method and the relevant simulation results in section~\ref{sec:simulation}. 
The theoretical model is introduced and analyzed in section~\ref{sec:theory}.
We summarize and conclude in section~\ref{sec:summary}.

\end{multicols}\twocolumn

\section{Monte Carlo simulations} 
\label{sec:simulation}

We have performed Monte Carlo simulations of a fluid of axially symmetric 
ellipsoidal particles with elongation 
$\kappa = \sigma_{\mbox{\tiny end-end}}/\sigma_{\mbox{\tiny side-side}}=3$.
The system was confined between two hard walls, to which chains of the
same particles were attached at one end. The grafting points 
were on a regular square lattice. The simulations were conducted in the 
$NPT$ ensemble, at constant temperature and pressure, and fixed number of 
particles. Two chain monomers and/or solvent particles with orientations 
$\uu_i$, $\uu_j$ ($|\uu| = 1$), whose centers are separated by a
distance vector $\rr_{ij}$, interact through the pair potential 
\begin{equation}
\label{eq:vij}
V_{ij}
= \left\{ \begin{array}{lcr}
4 \epsilon_0 \: (X_{ij}^{12} - X_{ij}^{6}) + \epsilon_0 & : & X_{ij}^6 > 
1/2 \\
0 & : & \mbox{otherwise}
\end{array} \right. .
\end{equation}
Here $X_{ij} = \sigma_0/(r_{ij}-\sigma_{ij}+\sigma_0)$ is an inverse 
reduced distance and the shape function~\cite{berne} 
\begin{eqnarray}
\lefteqn{\sigma_{ij}(\uu_i,\uu_j,\ru_{ij})
= \sigma_0 \:
\Big\{ 1 \: - \frac{\chi}{2}}  
\nonumber \\&& \quad
 \Big[
\frac{(\uu_i\cdot\ru_{ij}+\uu_j\cdot\ru_{ij})^2}
     {1+\chi \uu_i\cdot\uu_j}
+ 
\frac{(\uu_i\cdot\ru_{ij}-\uu_j\cdot\ru_{ij})^2}
     {1-\chi \uu_i\cdot \uu_j}
\Big] \Big\}^{-1/2}
\end{eqnarray}
approximates the contact distance in the direction of 
$\ru_{ij}=\rr_{ij}/r_{ij}$ of two ellipsoids with orientations 
$\uu_i$ and $\uu_j$~\cite{berne}. 
The anisotropy parameter $\chi$ is defined as $\chi=(\kappa^2-1)/(\kappa^2+1)$.
A hard core potential prevents the particles from penetrating the walls
at $z=0$ and $z=L_z$:
\begin{equation}
\label{eq:wall}
V_{W}(z) = \left\{ \begin{array}{l cc }
0 & :& \quad d_z(\theta) < z < L_z - d_z(\theta) \\
\infty & :& \quad \mbox{otherwise} 
\end{array} \right. ,
\end{equation}
\begin{displaymath} 
\mbox{with} \quad 
d_z(\theta) = \sigma_0/2 \: \sqrt{1+\cos^2 (\theta) \: (\kappa^2-1) }.
\end{displaymath}
The function $d_z(\theta)$ is the contact distance between the surface
and an ellipsoid of elongation $\kappa$ and diameter $\sigma_0$ oriented 
with an angle $\theta$ with respect to the surface normal.

Chain monomers are connected by bonds of length $b$, which are subject to 
a spring potential with an equilibrium length $b=b_0$ and a logarithmic 
cutoff at $|b-b_0|=b_s$.
\begin{equation}
\label{eq:fene}
V_{S}(b) = \left\{ \begin{array}{l c r}
\displaystyle - \frac{k_s}{2} b_s^2 \ln\Big( 1 - \frac{(b-b_0)^2}{b_s{}^2} \Big)
&: & |b-b_0| < b_s \\
\infty &: & |b-b_0| > b_s
\end{array} \right. .
\end{equation}
The bonds are made stiff and coupled to the orientations of the monomers 
by virtue of a stiffness potential
\begin{equation}
\label{eq:ba}
V_{A} (\theta_1,\theta_2,\theta_{12}) =
k_a \; [ \: 4 - \cos(\theta_1) - \cos(\theta_2) - 2 \cos(\theta_{12}) \: ],
\end{equation}
which depends on the angles $\theta_1$ and $\theta_2$ between 
the orientation of a monomer and the adjacent bonds, and the
angle $\theta_{12}$ between the two bonds.
The first bond of each chain is attached to one of the surfaces.

We chose the model parameters  $k_s=10 \epsilon_0/\sigma_0^2$,
$k_a=10 \epsilon_0$, $b_0=4.\sigma_0$ and $b_s=0.8\sigma_0$. 
The simulations were performed at the temperature $T=0.5 \epsilon_0/k_B$ 
and pressure $P=3 \epsilon_0/\sigma_0^3$. This corresponds to a 
state well in the nematic phase: The transition to the isotropic phase 
occurs at the pressure $P=2.3 \epsilon_0/\sigma_0^3$. 
The bulk number density was $\rho = 0.313/\sigma_0^3$.

We used a rectangular simulation box of size 
$L_{\parallel} \times L_{\parallel} \times L_z$ with periodic boundary 
conditions in the $x$ and $y$ direction and fixed boundary conditions 
in the $z$ direction. The Monte Carlo moves included particle
displacements, rescaling of the simulation box in the $z$ direction
(the lateral box size $L_{\parallel}$ was kept constant in order to
maintain a fixed grafting density), and a special variant of
configurational biased Monte Carlo moves~\cite{frenkel,ccp01}, in which chain
monomers are turned into solvent particles and new chains are grown
from the solvent. Details of this algorithm will be published
elsewhere~\cite{harald2,ccp01}. 

We have studied systems with roughly 2000 solvent particles (the numbers
varied slightly in the different runs) and up to 242 chains of four
monomers in simulation boxes of lateral size $L_{\parallel} = 12 \sigma_0$.
The length $L_z$ of the boxes fluctuated with $L_z \approx 65 \sigma_0$
at the highest grafting density.
The walls orient the solvent particles parallel to the surface. 
We started from an initial configuration where particles all pointed
into the $x$ direction. This was achieved by applying a strong orienting
field over roughly 50.000 Monte Carlo steps. After turning this field 
off, the system was equilibrated over at least 1 million Monte Carlo 
steps; data were then collected over 5 million or more Monte Carlo steps.

The results from these simulations for a range of grafting densities 
$\Sigma$ will be presented in detail elsewhere~\cite{harald3}.
Here, we focus on the anchoring effect that is the subject of this paper.
Figure~\ref{fig:snapshots} shows two configuration snapshots of systems 
at grafting densities $\Sigma = 0.34/\sigma_0^2$ and $\Sigma=0.84/\sigma_0^2$. 
The grafted chains are tilted in both cases. At the lower grafting 
density $\Sigma = 0.34/\sigma_0^2$, the tilt propagates into the bulk of the 
film. At $\Sigma = 0.84/\sigma_0^2$, the orientation of the liquid crystal 
outside in the bulk is perpendicular to the surface, even though 
the chains still retain tilt.

One might suspect an equilibration problem. In order to exclude this 
possibility, we have prepared an initial configuration in which all particles 
are oriented in the $z$ direction. It relaxes into the same structure 
as Fig. \ref{fig:snapshots}.

The effect can be characterized more quantitatively by inspection of
local order parameter profiles.  The order tensor of a system of $n$ 
particles is defined by~\cite{degennes}
\begin{equation}
\label{eq:op} 
\QQ = \frac{1}{n} \sum_{i=1}^{n} 
(\frac{3}{2} \uu_i \otimes \uu_i - \frac{1}{2} {\II} ) ,
\end{equation}
where $\uu_i$ denotes the orientation of the particle $i$ as above, 
$\II$ the unity matrix, and $\otimes$ the dyadic product. 

\begin{figure}[t]
\noindent
\hspace*{-0.5cm}
\fig{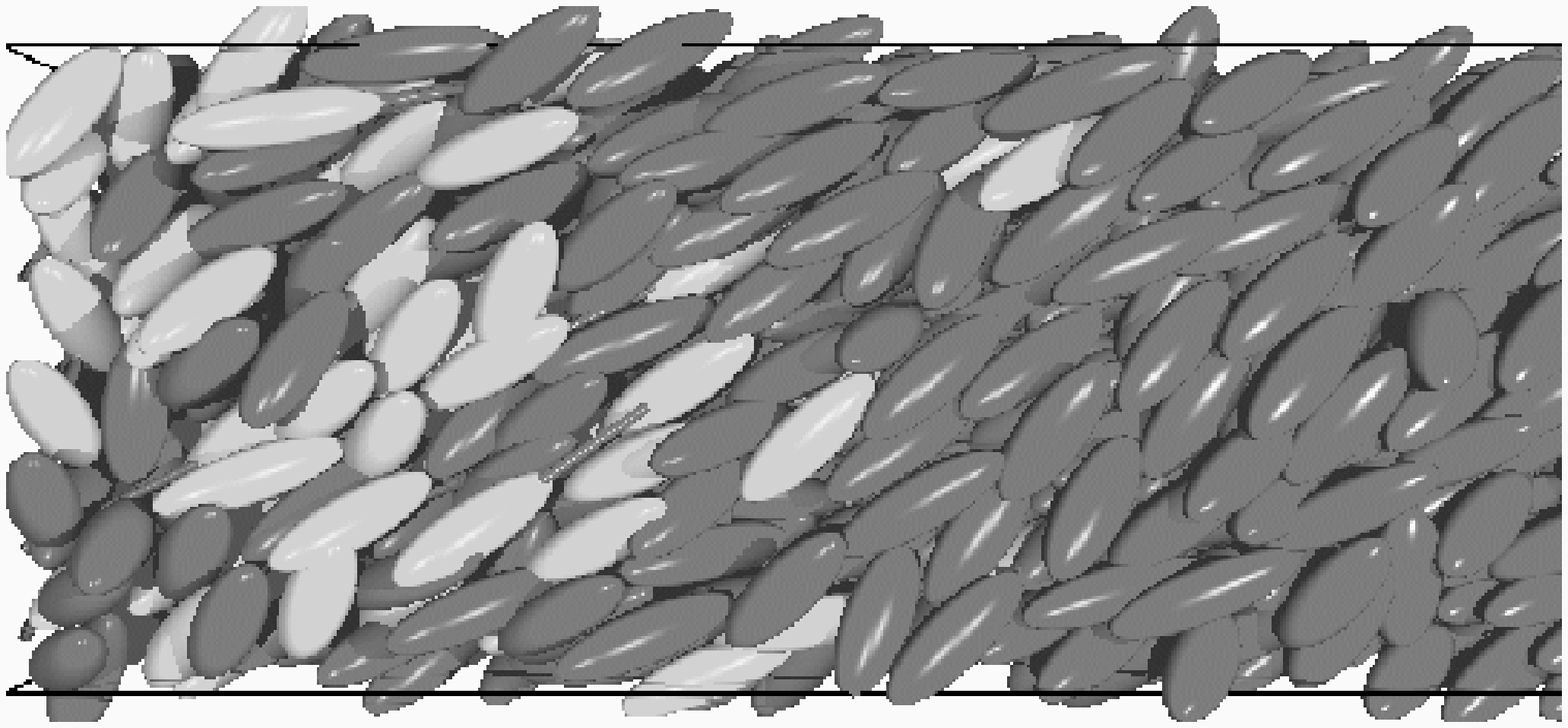}{80}{50}{
}\\
\hspace*{-0.5cm}
\fig{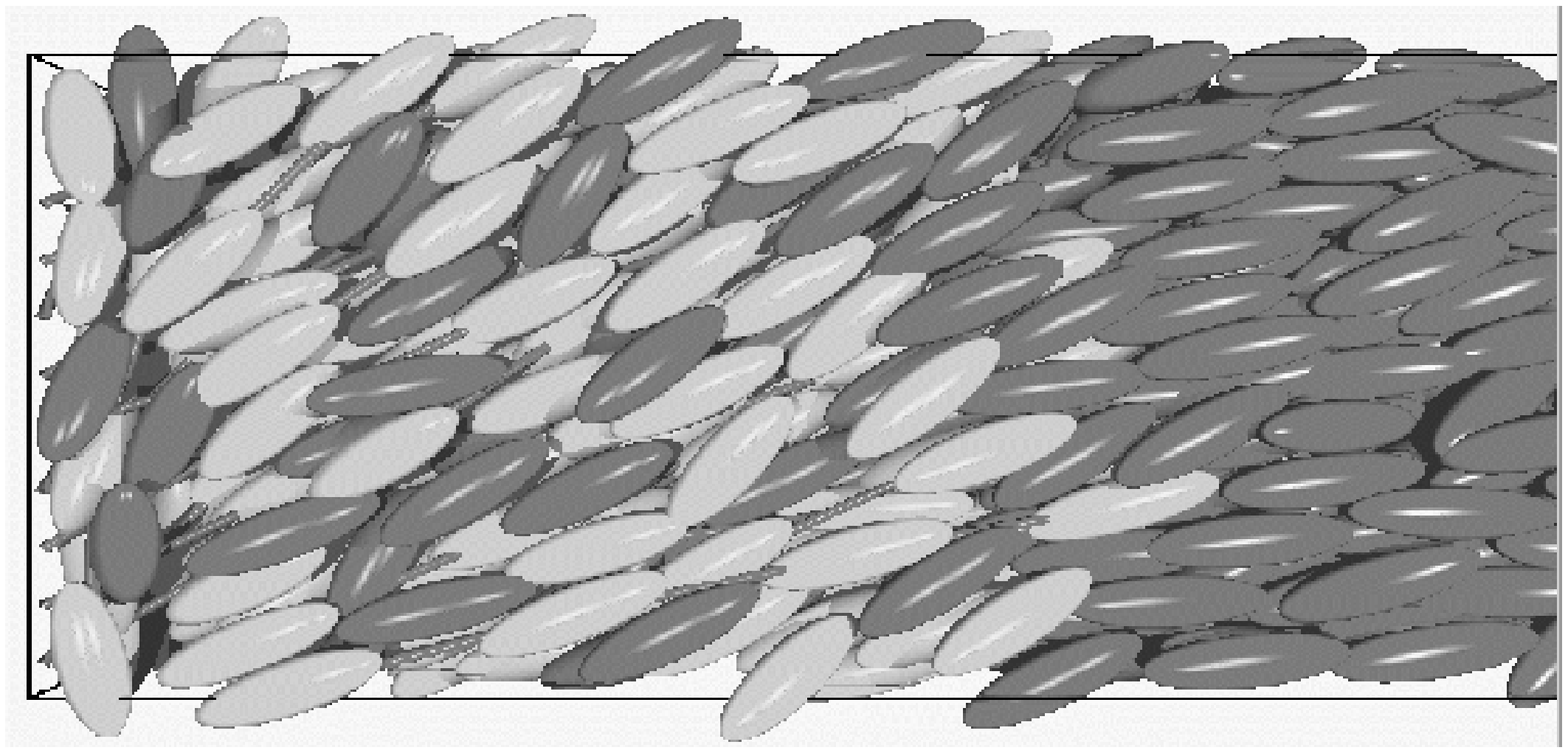}{80}{50}{
}
\vspace*{1cm}
\CCsnap
\end{figure}

\noindent
The largest eigenvalue of this matrix is the nematic order parameter $S$ 
(roughly 0.75 in the bulk of our system), and the corresponding eigenvector 
is the director $\nn$. In order to obtain profiles of these
quantities, we subdivide the system in the $z$ direction into slabs of 
thickness $\delta z = 0.1 \sigma_0$, and determine the order tensor in each 
slab. Note that the value of $S$ is larger in a slab than in the bulk 
due to the small number of particles in each slab. 

Figs. \ref{fig:profiles} a) and b) show averaged profiles of the order 
parameter and of the $z$-component of the director for the same grafting 
densities as in Fig. \ref{fig:snapshots}. The order parameter varies very 
little throughout the film; it is slightly larger in the chain region, 
due to fact that chain monomers have less rotational freedom than solvent
particles. The direction of alignment varies much more. Close to the 
surface, the particles are aligned parallel to the wall. 
However, the chains reorient the director within a few particle diameters 
from the surface. Then follows a region of slower changes, where the 
director gradually stands up. This region is restricted to the inside 
of the chain layer; further changes outside of the chain layer are small. 
In the case of $\Sigma = 0.84/\sigma_0^2$, the last monomers of the chains 
are almost perpendicular to the surface, even though the director is 
tilted inside of the chain region.

To characterize the chain layer further, Figs. \ref{fig:profiles} 
also show the density of chains and the density of chain ends (end monomers). 
A few chain ends are located very close to the substrate at distances 
$z = 1-2 \sigma_0$; they belong to chains which lie flat on the surface. 
The other chain ends are concentrated at the interface between the chain 
region and the bulk fluid. The higher the grafting density, the more
efficiently the chain ends are expelled from the inside of the chain 
layer towards the chain-bulk interface.

Since the cell width $L_{\parallel} = 12 \sigma_0$ is rather small,
quantitative details of the curves shown in Fig. \ref{fig:profiles} are 
presumably subject to finite size effects. The simulations were very time 
consuming, and we were not able to consider different system sizes. 
In the following, we shall focus on the qualitative observation that the 
chains produce homeotropic alignment beyond a certain grafting density, 
even though they themselves remain tilted.

\begin{figure}[t]
\noindent
\hspace*{-0.5cm}
\fig{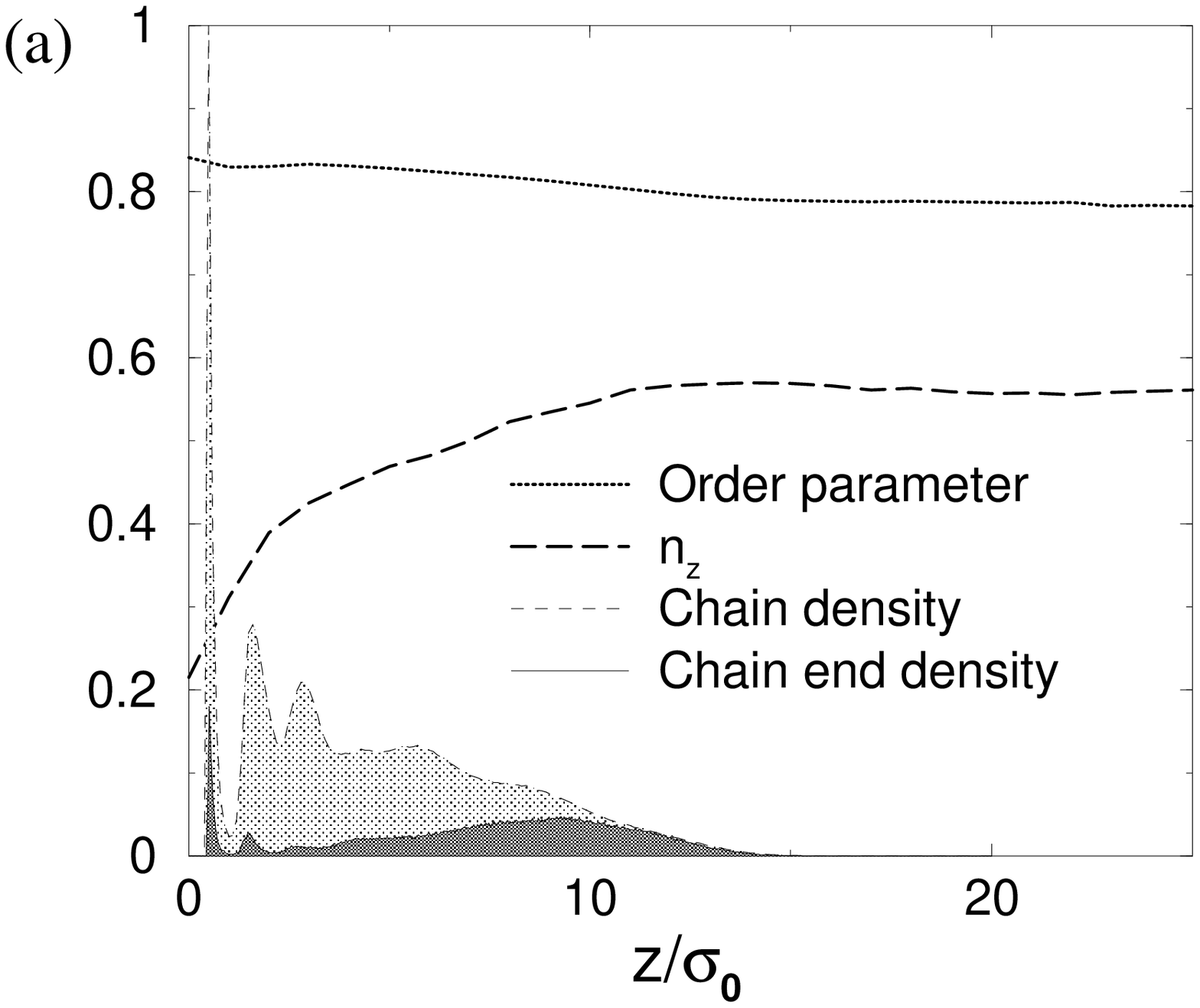}{70}{70}{
\vspace*{-0.5cm}
}\\
\hspace*{-0.5cm}
\fig{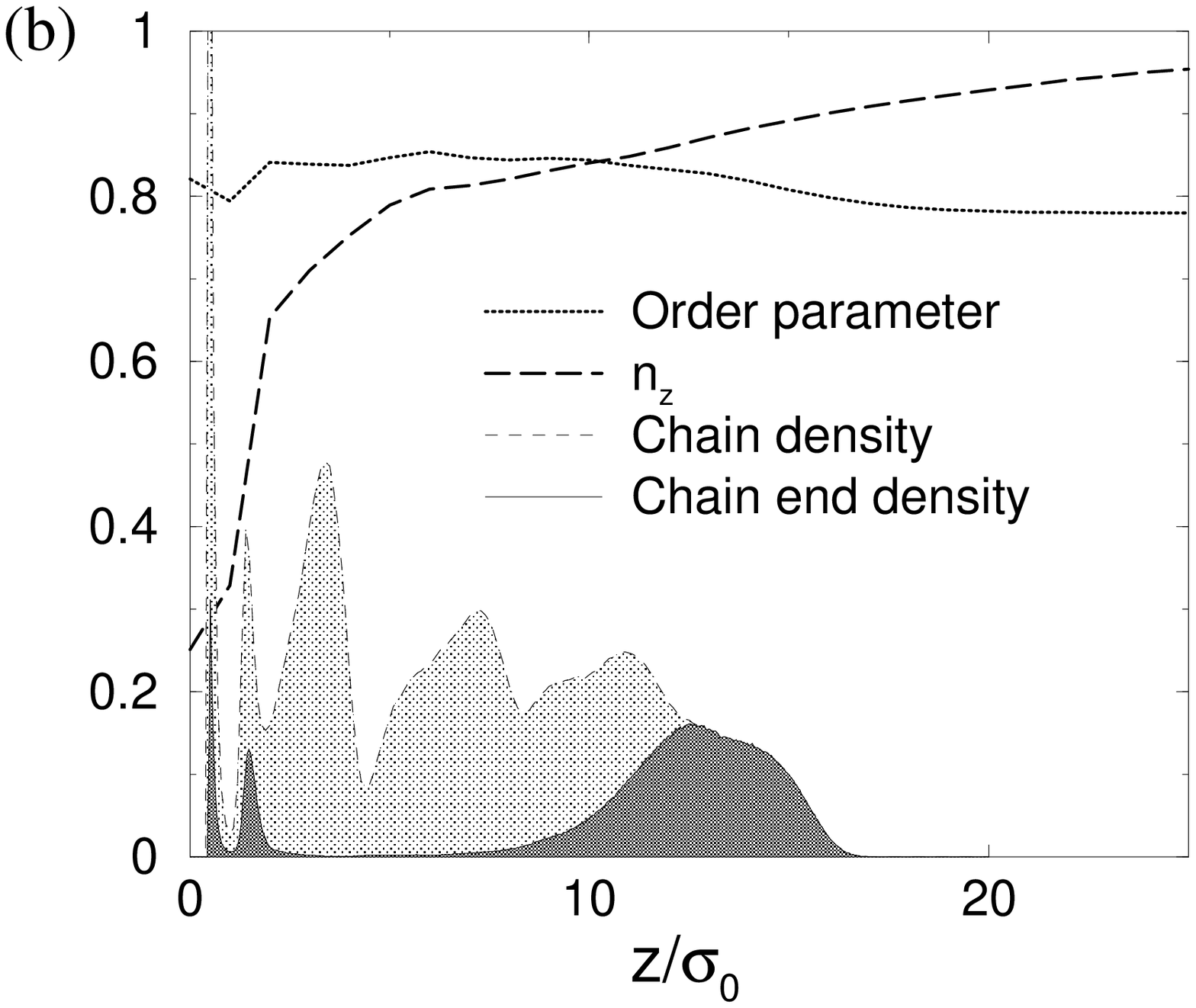}{70}{70}{
\vspace*{0cm}
\CCprofiles
}
\end{figure}

\section{A Theoretical Model} 
\label{sec:theory}

In this section, we shall propose a simple theoretical model for liquid
crystalline chains in a nematic solvent which can explain the phenomenon 
reported in the previous section. 

At first sight, the observation that the director orientation outside of
the chain layer differs from that in the chain layer is surprising.
The ``interface'' between the chain layer and the pure solvent seems to orient
the particles in a homeotropic way. However, the origin of that anchoring 
force is not obvious. If one admits the existence of such an aligning force, 
one is left with the question why the chains are still tilted. 
In the scenario of Halperin and Williams~\cite{avi1} the tilt is inherited 
from the bare substrate, which favors planar anchoring. In the simulations, 
the rapid director changes in the vicinity of the substrate suggest 
that the director orientation in the chain layer is largely decoupled from 
that at the substrate. Alternatively, collective tilt can be induced by 
attractive interactions between chains~\cite{me1}. Such interactions 
can be mediated by solvent particles due to local packing effects,
even if the direct interactions are purely repulsive~\cite{me2}.

Our model is designed to investigate on a simplified level the effect 
of such an interaction. We describe the chains by smooth differentiable 
paths $\RR(l)$ of length $L$ (wormlike chains~\cite{kratki,saito}), 
which follow exactly the director field of the surrounding solvent: 
$d\RR/dl = \nn(\RR)$, and are attached to the surface at one end. 
The grafting density is measured in terms of a dimensionless area 
fraction $\zeta$, which is defined as the grafting density times 
the cross section area of one chain. Fluctuations of the director 
are neglected. Moreover, we do not account for the possibility of 
hairpins~\cite{degennes2,warner,khodolenko}. 

The model has the following four key ingredients:
\begin{enumerate}
\item[1.] The elasticity of the nematic fluid.
\item[2.] The translational entropy of the solvent.
\item[3.] The interactions between the chains.
\item[4.] The anchoring energy of the substrate.
\end{enumerate}

The elastic energy is given by the Frank free energy 
functional~\cite{oseen,zocher,frank}
\begin{eqnarray}
\lefteqn{{\cal F}_{\mbox{\tiny Frank}} \{ \nn (\rr) \} 
= \frac{1}{2}\int d \rr \Big\{ K_{11} [ \nabla \cdot \nn ]^2 +} \qquad
\nonumber \\ &&
\label{eq:frank}
K_{22} [ \nn \cdot (\nabla \times \nn) ]^2 +
K_{33} [ \nn \times (\nabla \times \nn) ]^2 \Big\},
\end{eqnarray}
with the Frank elastic constants $K_{11}$ (splay), $K_{22}$ (twist)
and $K_{33}$ (bend). 

The translational entropy of the solvent is calculated within
the Flory-Huggins approximation~\cite{flory1,huggins,flory2}
\begin{equation}
\label{eq:solvent}
{\cal F}_{\mbox{\tiny solvent}} \{ \Phi(\rr) \}
= \frac{k_B T}{v_s} \int d \rr \: (1 - \Phi) \ln (1 - \Phi).
\end{equation}
Here $k_B$ is the Boltzmann factor, $T$ the temperature, 
$v_S$ the volume occupied by a solvent particle, and
$\Phi(\rr)$ the local volume fraction of chain monomers.

The excluded volume interactions between chains are incorporated
by the constraint that $\Phi(\rr)$ must never exceed one. 
The attractive interactions are described by a quadratic free energy
contribution,
\begin{equation}
\label{eq:ww_bulk}
{\cal F}_{\mbox{\tiny chains}} \{ \Phi(\rr) \} 
=  - \frac{v}{2} \int d \rr \Phi^2,
\end{equation}
where $v$ is an effective interaction strength. Since the chains are
monodisperse and have no hairpins, their ends are located on a smooth, 
well-defined surface. In the vicinity of that surface, the number of
interacting neighbors of a monomer is reduced, which in turn reduces
the interaction energy. We account for this by adding a 
correction term
\begin{equation}
\label{eq:ww_surf}
{\cal F}_{\mbox{\tiny surface}} \{ \Phi(\rr); \nn(\rr) \}
= v \int_{\mbox{\tiny surface}} \!\!
 d {\bf s}  \: \Phi^2 \: \gamma(\Phi,\nn,\uu_{{\bf s}}).
\end{equation}
The integral $\int_{\mbox{\tiny surface}} \!\!\! d {\bf s} $ runs over 
the surface of the chain layer, and $\uu_{{\bf s}}$ denotes the unit vector
in direction of the surface normal. 
The function $\gamma$ integrates over the fraction of ``missing''
interaction energy in a region close to the surface. 
We take $\gamma$ to be proportional to the thickness of the affected 
region, which we calculate as illustrated in Figure \ref{fig:cartoon1}: 
$\gamma \propto a \sin (\theta)$,
where  $\theta$ is the angle between the monomer orientation $\nn$ 
and the surface normal $\uu_{{\bf s}}$, and $a$ the distance between
two neighbor chains, $a \propto 1/\sqrt{\Phi}$. This yields
\begin{equation}
\label{eq:gamma}
\gamma(\Phi,\nn,\uu_{{\bf s}}) = \rg \: \Phi^{-1/2} \sqrt{1-(\nn \cdot \uu)^2}.
\end{equation}
The proportionality constant $\rg$ depends on the form of the chain 
interactions and the actual chain packing at the surface, and
is presumably of order 1. 

\begin{figure}[t]
\noindent
\hspace*{0.cm}
\fig{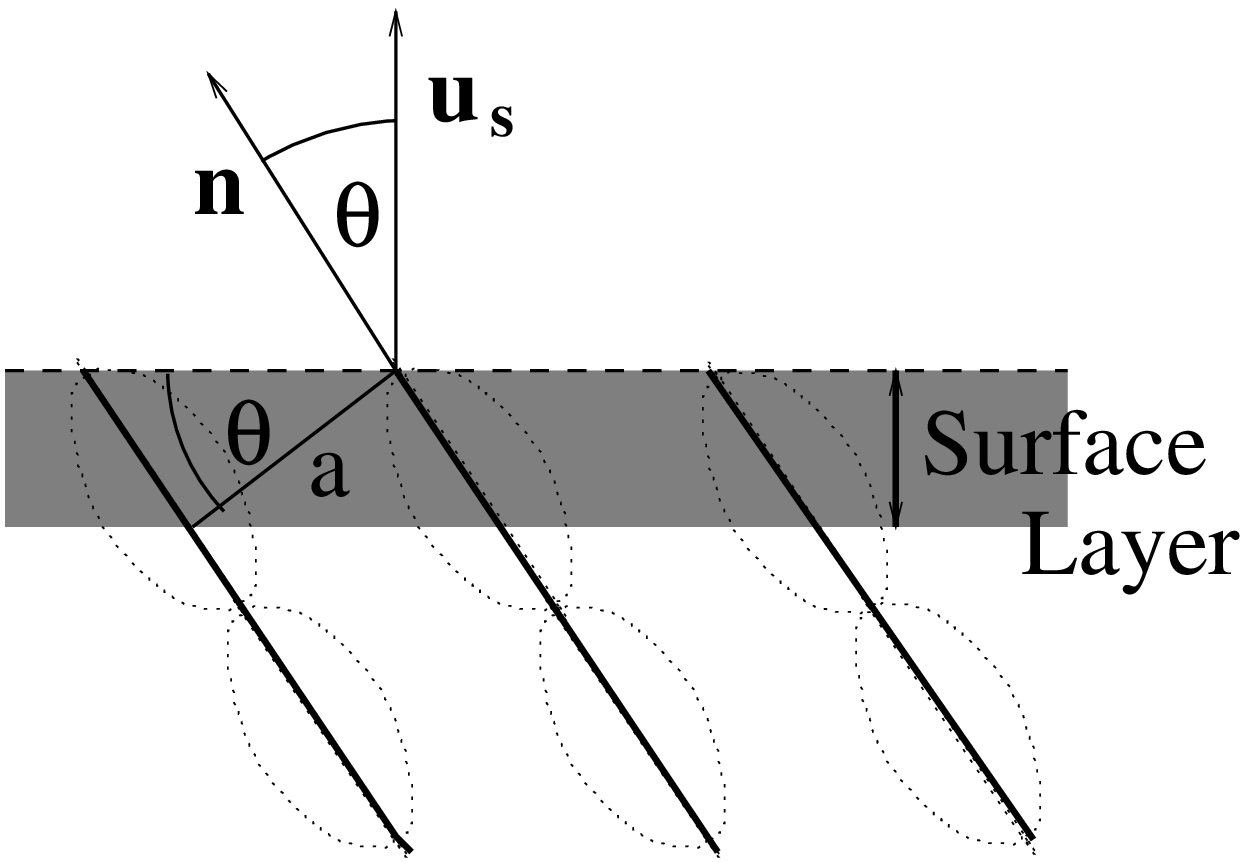}{60}{45}{
\vspace*{0.8cm}
\CCcartoon
}
\end{figure}
\noindent

The free energy contribution (\ref{eq:ww_surf}), (\ref{eq:gamma}) turns 
out to be essential for the effect that we wish to study. On the other hand, 
the considerations which led to eqn. (\ref{eq:gamma}) may seem somewhat 
crude and oversimplified. Therefore, we add some comments which justify 
an ansatz of the form (\ref{eq:gamma}) on more general grounds: 
In an anisotropic system, one clearly expects the missing neighbor 
contribution (\ref{eq:ww_surf}) at the surface of the chain layer to be 
anisotropic, i.e., to depend on the angle $\theta$ between the surface normal 
and the director. Since the chains are aligned along the director, 
interchain interactions take place in directions perpendicular to the 
director mainly. Hence fewer interactions are missing if the director 
is perpendicular to the surface, and the missing neighbor contribution 
should favor small $\theta$.  The ansatz (\ref{eq:gamma}), 
${\cal F}_{\mbox{\tiny surface}} \propto \sin(\theta)$, 
ensures this in the simplest possible way. Other expressions which 
favor small $\theta$ lead to similar results.

Finally, the orienting effect of the bare substrate is described by
a quadratic potential
\begin{equation}
\label{eq:substrate}
{\cal F}_{\mbox{\tiny substrate}} \{ \nn(\rr) \}
 = W \int_{\mbox{\tiny substrate}} \!\!
  d {\bf s} \: (\nn \cdot \uu_{{\bf s}})^2.
\end{equation}
The surface integral $\int \! d{\bf s}$ runs over the surface of the substrate. 
Other substrate-related free energy contributions, such as missing 
neighbor effects of the type (\ref{eq:ww_surf}) and solvent mediated 
interactions between the chains and the substrate are taken to be negligible 
compared to the anchoring energy (\ref{eq:substrate}). Furthermore, we neglect
the internal stiffness of chains, and disregard the possibility that 
the elastic constants may vary locally and depend on the volume fraction 
occupied by the chains. These contributions can be incorporated in our model 
in a relatively straightforward way, but at the expense of having more model 
parameters. The results do not change qualitatively.

Since fluctuations are neglected, quantities vary only in the direction 
perpendicular to the surface, the $z$-direction, and the director $\nn$ 
does not rotate in the $xy$-plane. Inside of the chain layer, the local volume 
fraction $\Phi(z)$ of the chains is given by $\Phi(z) = \zeta/n_z(z)$. 
Outside of the chain layer, $\Phi(z)$ is zero and the director is constant. 
Hence the system can be described entirely in terms of the profile $n_z(z)$ 
inside of the chain layer. The free energy per surface area is given by
\begin{eqnarray}
{\cal F}\{n_z(z)\} &=&
\int_0^H dz \:
\Big\{ \: \frac{1}{2}(\frac{d n_z}{dz})^2 
(K_{11} + K_{33} \frac{n_z{}^2}{1-n_z{}^2}) 
\nonumber \\ &&
+ \: \frac{k_B T}{v_s} (1-\frac{\zeta}{n_z}) \ln(1-\frac{\zeta}{n_z})
 - \frac{v}{2} (\frac{\zeta}{n_z})^2 \: \Big\}
\nonumber \\ &&
+ \: W n_z^2|_{z=0} 
+ \rg v \sqrt{\zeta^3} \sqrt{\frac{1-n_z{}^2}{n_z{}^3}}|_{z=H}.
\label{eq:f_nz}
\end{eqnarray}
The height $H$ of the chain layer is determined implicitly by the requirement
$L = \int_0^H dz/n_z(z)$. 

For practical calculations, it is convenient to perform the variable 
transformation $dl = dz/n_z(z)$, i.e., to consider variations along 
the chains rather than variations in the $z$-direction.
Furthermore, we define the units of length and energy~\cite{footnote1} 
such that $(K_{11}+K_{33})/2 = 1$ and $k_B T/v_s = 1$, and introduce 
the notation $\delta  K = (K_{33}-K_{11})/(K_{33}+K_{11})$.
The free energy per surface area is then expressed as a functional of the 
profile $\theta(l) = \arccos(n_z(l))$ of the angle $\theta$ between $\nn$ and
the surface normal:
\begin{eqnarray}
{\cal F}\{\theta(l)\} &=&
\int_0^L dl \:
\Big\{ \: \frac{1}{2} C(l) (\frac{d \theta}{dl})^2  + U(\theta) \: \Big\}
\nonumber \\ &&
+ \: W \: \cos^2(\theta) |_{l=0} 
+ \rg v \sqrt{\zeta^3} 
   \frac{\sin(\theta)}{\sqrt{\cos^3(\theta)}}|_{l=L},
\label{eq:f_theta}
\end{eqnarray}
with
\begin{eqnarray}
\label{eq:c}
 C(\theta) &=& \frac{1}{\cos(\theta)} \: (1 + \delta K \: \cos(2 \theta)) \\
\label{eq:u}
 U(\theta) &=& \zeta \hat{U}(\frac{\cos(\theta)}{\zeta})  
\end{eqnarray}
\begin{displaymath}
\mbox{and} \qquad \hat{U}(x) = (x-1) \ln (1-\frac{1}{x}) - \frac{v}{2x}.
\end{displaymath}
Minimizing this functional yields the Euler-Lagrange equation
\begin{equation}
\label{eq:el}
C \frac{d^2 \theta}{dl^2} + \frac{1}{2} (\frac{d \theta}{dl})^2
\frac{d C}{d \theta} - \frac{dU}{d \theta} = 0.
\end{equation}
and the boundary conditions
\begin{eqnarray}
\label{eq:bl0}
l=0&:&  \qquad
C(\theta) \frac{d \theta}{dl} = -2 W \cos(\theta) \sin(\theta) \nonumber \\
&& \qquad \mbox{or} \quad \theta = \arccos(\zeta) \\
\label{eq:bll}
l=L&:& \qquad
C(\theta) \frac{d \theta}{dl} = -\rg \: v \sqrt{\zeta^3}
\frac{1+3/2 \: \tan^2(\theta)}{\sqrt{\cos(\theta)}} \nonumber\\
&& \qquad \mbox{or} \quad \theta = 0.
\end{eqnarray}
Eqn.~(\ref{eq:el}) describes the trajectory of a particle in space 
$\theta$ and time $l$ with a position dependent mass $C(\theta)$, 
which moves in the potential $-U(\theta)$. The boundary conditions
pull the particle in the direction of large $\theta$ at $l=L$ and
in the direction of small $\theta$ at $l=0$. Hence the profile of
$\theta$ decreases monotonically with $l$. The anchoring angle on 
the chain layer is given by $\theta(L)$.

The properties of the model are most easily discussed in the limit
of long chains, $L \to \infty$. The two boundaries of the chain layer at 
$l=0$ and $l=L$ are then decoupled, and there exists a region inside of 
the chain layer in which $\theta(l)$ is constant, $\theta(l) \equiv \bt $. 
The stationary angle $\bt$ is the angle where $U(\theta)$ has its minimum. 
It depends on the grafting density $\zeta$ and the chain interaction parameter 
$v$. If $v>1$, the chains exhibits a continuous transition from an untilted 
state at large $\zeta$ to a tilted state at $\zeta < \zeta_c(v)$, where 
$\zeta_c(v)$ is determined by the implicit equation
\begin{equation}
\label{eq:vc}
v = - \frac{2}{\zeta_c(v)^2} \{\zeta_c(v) + \ln [1-\zeta_c(v)]\}.
\end{equation}
In the vicinity of the phase transition, at $\zeta \approx \zeta_c(v)$, 
the potential $U(\theta)$ can be expanded like
\begin{equation}
\label{eq:u_c}
U(\theta) = 
\mbox{const.} - \frac{1}{2} (\zeta_c - \zeta) \: b(\zeta_c) \:\theta^2
+ \frac{1}{8} \zeta_c \: b(\zeta_c) \: \theta^4
\end{equation}
with $b(\zeta) = (2 - \zeta)/(1 - \zeta) + 2/\zeta \: \ln(1-\zeta)$,
and the minimum is found at 
\begin{equation}
\label{eq:t_c}
\bt = \sqrt{2 (\zeta_c - \zeta)/\zeta_c}.
\end{equation}
A necessary criterion for the validity of the expansion is obviously
$\Phi = \zeta/\cos(\bt) < 1$, which is fulfilled if
\begin{equation}
\label{eq:crit}
\zeta_c - \zeta < 2 \zeta_c \arccos^2(\zeta_c).
\end{equation}
Far from the transition, $U(\theta)$ takes its minimum close to the 
maximum value of $\theta$ at $\cos(\bt)/\zeta \approx 1$ and one obtains
\begin{equation}
\label{eq:u_f}
U(0) - U(\bt) = (1-\zeta)[\ln (1-\zeta) +  v \zeta/2].
\end{equation}

In the case $L \to \infty$, the anchoring angle $\theta(L)$ of the nematic 
film at the surface of the chain layer does not depend on the anchoring 
strength $W$ of the substrate. Exploiting the integration constant of the 
Euler-Lagrange equation (\ref{eq:el}),
$U(\theta) - \frac{1}{2} C(\theta) (\frac{d \theta}{d l})^2 \equiv$ const. 
$= U(\bt)$, one can rewrite the boundary condition (\ref{eq:bll}) at $l=L$ 
as
\begin{equation}
\label{eq:bll_2}
U(\theta(L))-U(\bt) = g(\theta)
\qquad \mbox{or} \qquad \theta(L) = 0.
\end{equation}
\begin{displaymath}
\mbox{with} \quad
g(\theta) = 2 \rg^2 v^2 \zeta^3 
\frac{ (1 + 3/2 \: \tan^2(\theta))^2}{1 + \delta K \: \cos(2\theta)}
\end{displaymath}

\begin{figure}[t]
\noindent
\hspace*{0.cm}
\fig{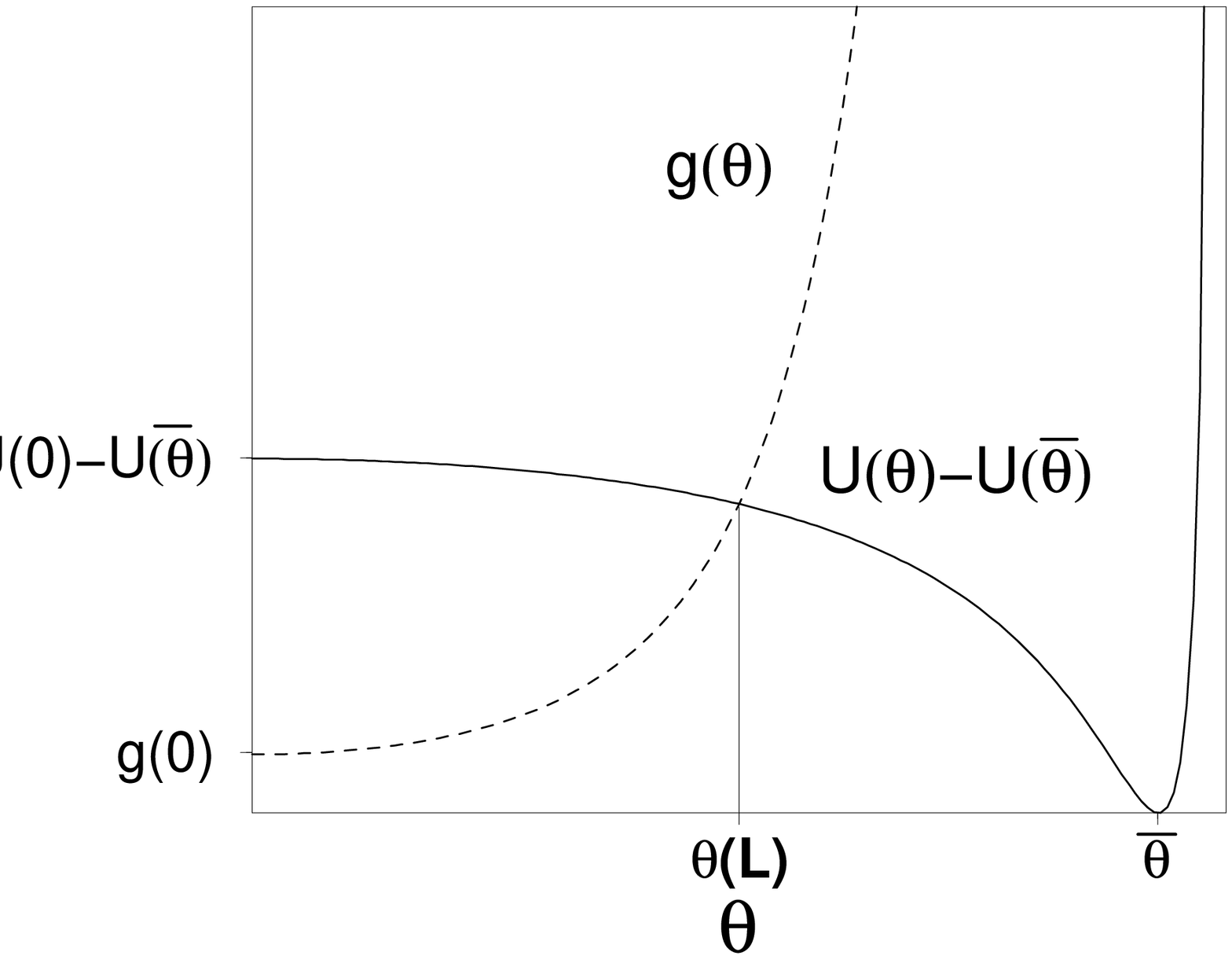}{65}{68}{
\vspace*{0cm}
\CCboundary
}
\end{figure}

The problem can be solved graphically as sketched in Fig.~\ref{fig:boundary}.
Since $g(\theta)$ grows monotonically for all $\delta K < 1$, the
curves $g(\theta)$ and $U(\theta)-U(\bt)$ cross at most at one 
point in the region $\theta < \bt$. If such a point exists,
it defines the anchoring angle $\theta(L)$. Otherwise, the case
$\theta(L) = 0$ 
applies and the anchoring is homeotropic. 
The transition between homeotropic and tilted anchoring is continuous.
The grafting density at the transition $\zeta^*(v)$ can be calculated by 
solving $U(0)-U(\bt)=g(0)$ for a given interaction parameter $v$. If $v$ is 
small, $\zeta^*(v)$ is close to the tilting transition $\zeta_c(v)$ of the 
chains, and Eqn.~(\ref{eq:t_c}) applies. The tilt angle $\bt^*$ of the chains
at the transition can be approximated by its asymptotic value at small 
grafting densities $\zeta^*(v) \stackrel{v \to 1}{\longrightarrow} 0$. 
Using Eqns.~(\ref{eq:vc}) and expanding $\cos(\bt^*)$ in
powers of $\bt^*$, one obtains after some algebra
\begin{equation}
\label{eq:low}
\zeta^*(v) \sim \zeta_c(v) / (1+ 2 \rg \sqrt{3/(1 + \delta K)}).
\end{equation}
At large $v$, $\zeta^*$ is far from $\zeta_c$ and Eqn.~(\ref{eq:u_f}) leads to
\begin{equation}
\label{eq:high}
\zeta^*(v) \sim \sqrt{1 + \delta K}/(2 \rg \sqrt{v}).
\end{equation}

Fig.~\ref{fig:phdiag} shows two examples of phase diagrams for the
parameter values $\rg/\sqrt{1 + \delta K} = 0.5$ and $0.1$. 
We expect that typical values of $\rg/\sqrt{1 + \delta K}$ are in
that range. At high grafting density $\zeta$ and low chain interaction
parameter $v$, the chains stand upright, perpendicular
to the surface. The thin line marks the continuous transition
$\zeta_c(v)$ to a phase where they tilt collectively. However, the 
orientation of the director outside of the chain layer still remains 
perpendicular. Only at grafting densities below a second critical
value $\zeta^*(v) < \zeta_c(v)$ (thick line) do the chains 
induce tilted alignment in the bulk of the nematic fluid.

In agreement with the observations from computer simulations reported 
in the previous section \ref{sec:simulation}, we thus find a region 
in $\zeta-v$ space where the chains are tilted, but nevertheless align the 
solvent in a homeotropic way. The size of that region increases 
with $\rg$, i.e., with the amplitude of the (anisotropic) missing neighbor 
effect at the surface of the chain layer. It is bounded by two continuous 
phase transitions - one at high grafting density to a phase where both the 
chains and the adjacent fluid are untilted, and one at low grafting
density to a phase where they are both tilted.

So far we have discussed this phenomenon for the limit of long chains, 
$L \to \infty$. In systems with shorter chains, the scenario remains 
qualitatively similar. If the anchoring on the bare substrate is planar 
($W > 0$ in Eqn.~(\ref{eq:substrate})), the sharp tilting transition 
in the chain region is replaced by a smoother crossover from distinctly 
tilted to roughly perpendicular chain orientation~\cite{footnote2}, and 
the anchoring transition shifts to higher grafting densities. One still 
finds a region where markedly tilted chains align the nematic solvent 
in a homeotropic way.

\begin{figure}[t]
\noindent
\fig{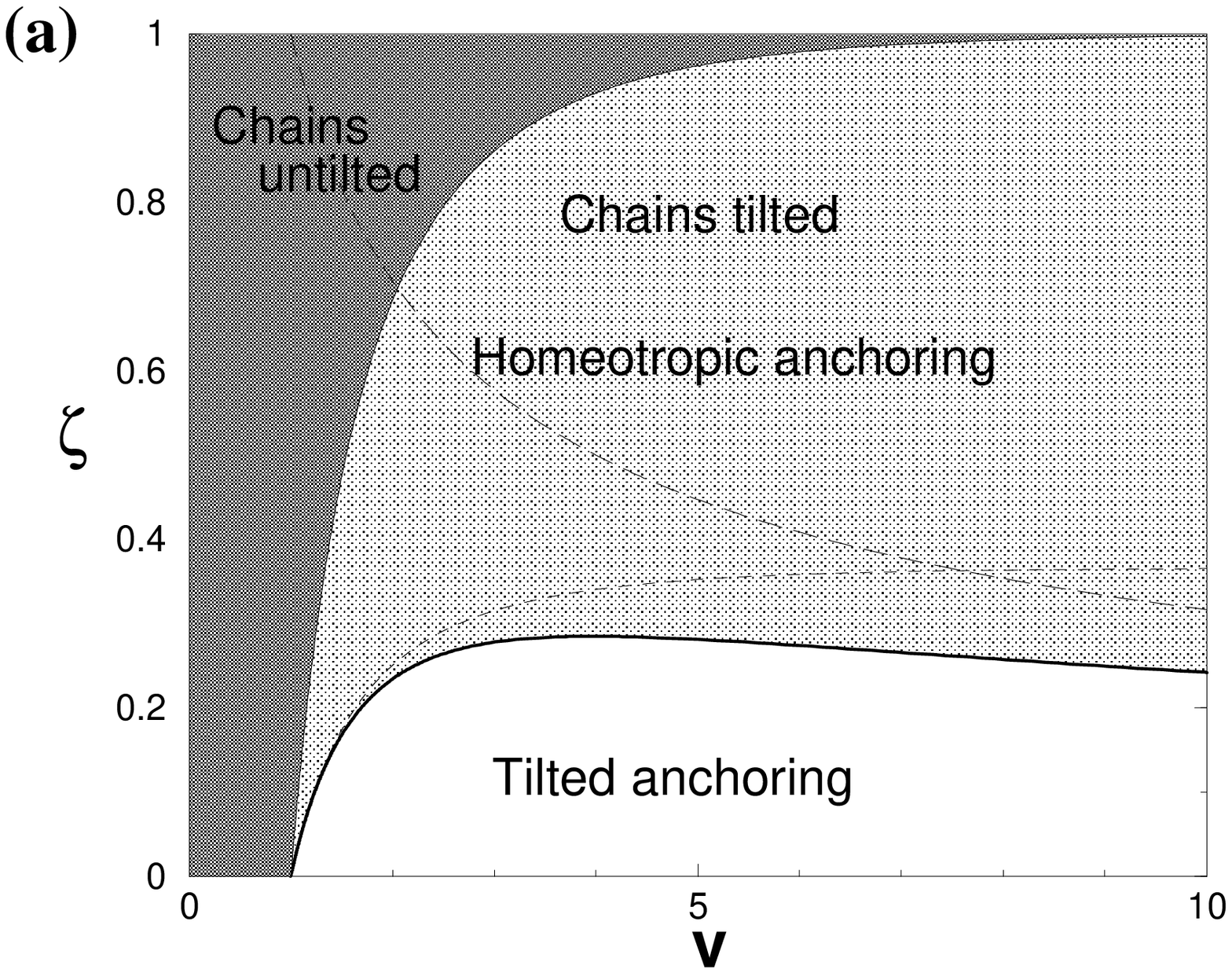}{70}{65}{
}\\
\fig{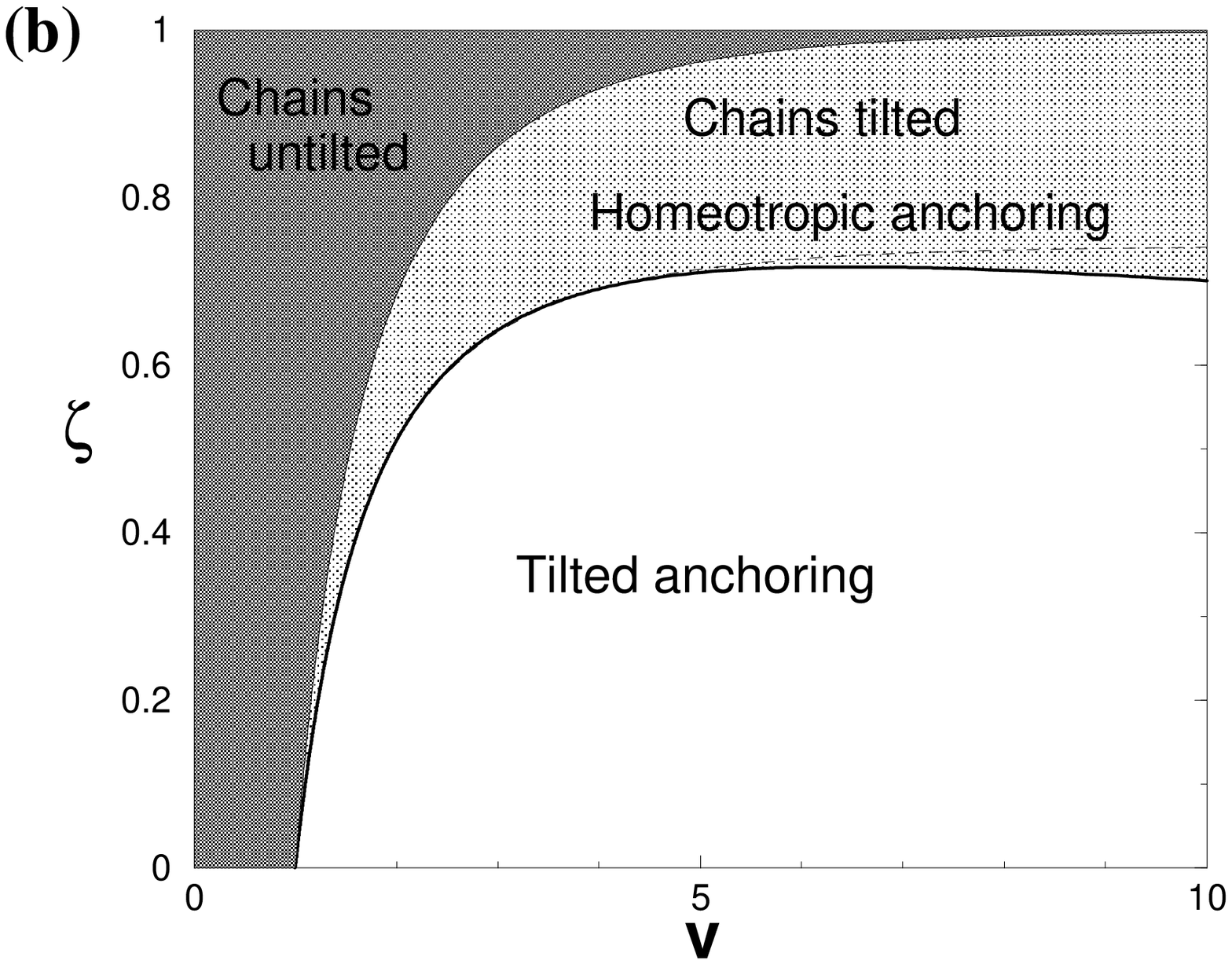}{70}{63}{
\vspace*{0cm}
\CCphdiag
}
\end{figure}

\section{Summary and Discussion}
\label{sec:summary}

To summarize, we have described a new type of anchoring transition
on grafted layers of main-chain liquid-crystalline chain molecules. 
It is driven by the competition between attractive chain interactions,
the translational entropy of the nematic solvent in a region close to 
the surface of the chain layer, and the elastic energy of the nematic fluid
which opposes rapid director changes. Inside of the chain layer, the 
chain interactions make the chains tilt collectively. Close to the surface, 
the interactions are reduced, and an increased number of solvent particles 
enters and swells the chains. As a result, chain ends in the surface region 
stand up. The anchoring angle imposed by the chains on the adjoining nematic 
fluid is thus smaller than the actual tilt angle of the chains. Above a 
critical grafting density $\zeta^*$, it drops to zero, despite the fact that 
the chains are still tilted. We have observed this effect in computer 
simulations of a model system with swollen grafted chains, and rationalized it 
in terms of a simple theoretical model.

The new anchoring transition is controlled by the grafting density 
of the chains. In that sense, it is similar to that discussed by 
Halperin and Williams~\cite{avi1,avi2,avi3}. However, the 
phase transition and the underlying mechanism is very different. 
In the case studied by Halperin and Williams, the chains are long and 
flexible enough to support many hairpins. The solvent is taken to be 
very good, i. e., chain monomers effectively repel each other. 
The transition is driven by an interplay between the conformational 
entropy of the chains, the anchoring force of the substrate on 
the solvent, and the elastic energy in the brush. The transition
connects a phase with planar anchoring and one with tilted anchoring. 

In the present paper, we consider a different regime: The
chains are shorter or stiffer, they have no hairpins, and their 
conformational entropy is negligible. The solvent is not good, chains 
effectively attract each other and there exists a well-defined interface 
between the chain layer and the solvent. The anchoring behavior is controlled 
by the structure of that interface. The influence of the bare substrate 
is minor. The transition connects a phase with tilted anchoring
and one with perpendicular anchoring. 

One might ask whether there could be a second transition between planar 
and tilted anchoring, at a lower grafting density, which would resemble 
that discussed by Halperin and Williams. Our theoretical model does not 
predict such a transition. In the simulations, the situation is more 
complicated. The data indicate the presence of a first order transition,
which has nothing to do with hairpins, but is still related to the 
organization of the chains. This will be described in 
detail in a separate publication~\cite{harald2,harald3}.

Experimentally, one finds that liquid crystalline polymers are
not swollen very well by nematic solvents~\cite{peng3}. 
We thus believe that our effect can be observed in real systems. 
The grafting density must be high to achieve low anchoring angles 
$\theta$, on the other hand, the chains need not be much longer
than their persistence length. Moreover, our results indicate that the 
grafting density required for the transition decreases with decreasing 
solvent quality.  Brushes with high grafting densities can be prepared
with the ``grafting from'' technique developed by R\"uhe et 
al~\cite{prucker}. This technique has been used in the experiments 
of Peng et al~\cite{peng1,peng2} mentioned in the introduction. 
However, Peng et al studied brushes with liquid crystalline 
side chains, where the mesogenic units are preferably oriented 
perpendicular to the backbone of the chains. The swelling of the 
brush in the surface region thus promotes planar anchoring, 
and no transition to homeotropic anchoring can be expected. 
The effect discussed in the present paper will presumably transpire 
in a similar experiment with short main-chain liquid-crystalline polymers.

One necessary requirement is of course that the orientation
of the chains is homogeneous. This turned out
to be a problem experimentally~\cite{peng1}, the brushes tend to 
exhibit planar multidomains. However, recent work indicates that 
an appropriate treatment of the bare substrate before growing 
the brush can force the chains into one monodomain~\cite{peng2}.
If brushes could be grown with grafting densities close to the
transition density $\zeta^*$, it would be possible to prepare
surfaces which anchor the solvent at any given small tilt
angle $\theta^*$. This would potentially have applications
in the design of liquid crystal display devices. 

Future theoretical work will have to study in more detail the chain
length dependence. The theory presented here has been devised for 
long chains -- for example, they have been described by homogeneous 
strings -- but the chains in the simulations were very short with only 
four monomer units. Chains in real brushes are usually polydisperse.
The influence of polydispersity on the anchoring behavior is not clear. 
Thus far the analytical model calculations have disregarded the 
possibility of lateral structure and lateral fluctuations. 
These will certainly become important for polydisperse brushes 
and more generally at low grafting densities.

\section*{Acknowledgments}
 
We thank M. P. Allen and K. Binder for useful discussions, and 
for allowing us to perform a major part of the simulations on the 
computers of the university of Bristol and the university of Mainz. 
We have also benefitted from stimulating interactions with 
D. Johannsmann, J. R\"uhe and A. Halperin. This work was funded
by the German Science Foundation (DFG).


\begin{thebibliography}{99}
\bibitem{degennes} P.-G. de Gennes and J. Prost, 
  {\em The Physics of Liquid Crystals} 
  (Oxford University Press, Oxford, 1995).
\bibitem{chandrasekhar} S. Chandrasekhar,
  {\em Liquid Crystals} 
  (Cambridge University Press, Cambridge, 1992).
\bibitem{jerome} B. Jerome, 
  Rep. Progr. Phys. {\bf 54}, 391 (1991).
\bibitem{bahadur} B. Bahadur (edt.) 
  {\em Liquid crystals and uses}, World Scientific, Singapore (1990).
\bibitem{schadt} M. Schadt, 
  Ann. Rev. Mater. Science, {\bf 27}, 305 (1997).
\bibitem{toney} M.F. Toney, T.P. Russell, J.A. Logan, H. Kiguchi, 
  J.M. Sands, S K. Kumar,
  Nature, {\bf 374}, 709 (1995).
\bibitem{abbott} N. L. Abbott, 
  Curr. Opn. Coll. Interf. Science {\bf 2}, 76 (1997).
\bibitem{avi1} A. Halperin, D. R. M. Williams,
  Europhys. Lett. {\bf 21}, 575 (1993).
\bibitem{avi2} A. Halperin, D. R. M. Williams,
  J. Physics: Cond. Matt. {\bf 6}, A297 (1994).
\bibitem{avi3} A. Halperin, D. R. M. Williams,
  Ann. Rev. of Mat. Science {\bf 26}, 279 (1996).
\bibitem{peng1} B. Peng, D. Johannsmann, J. R\"uhe,
  Macromolecules {\bf 32}, 6759 (1999).
\bibitem{peng2} B. Peng, J. R\"uhe, D. Johannsmann,
  Adv. Mater. {\bf 12}, 821 (2000).
\bibitem{prucker} O. Prucker, J. R\"uhe, 
  Macromolecules {\bf 31}, 592 (1998); {\em ibid}, 602 (1998).
\bibitem{peng3} F. Benmouna, B. Peng, J. R\"uhe, D. Johannsmann,
  Liqu. Cryst. {\bf 26}, 1655 (1999).
\bibitem{me1} F. Schmid, D. Johannsmann, A. Halperin, 
  J. de Physique {\bf 6}, 1331 (1996). 
\bibitem{berne} B. J. Berne and P. Pechukas,
 J. Chem. Phys. {\bf 56}, 4213 (1975).
\bibitem{frenkel} D. Frenkel, B. Smit,
  {\em Understanding Molecular Simulation}, 
  Academic Press, San Diego (1996).
\bibitem{harald2} H. Lange, 
  Dissertation Universit\"at Mainz (2001).
\bibitem{ccp01} H. Lange, F. Schmid,
  submitted to Comp. Phys. Comm. (2001).
\bibitem{harald3} H. Lange, F. Schmid,
  manuscript in preparation.
\bibitem{me2} M. M\"uller, F. Schmid, in
  {\em Annual Reviews in Computational Physics} VI, pp. 59, 
  D. Stauffer edt., World Scientific, Singapore (1999). 
\bibitem{kratki} O. Kratki, G. Porod, 
  Recl. Trav. Chim. 68, 1106 (1949).
\bibitem{saito} N. Saito, K. Takahashi, Y. Yuniki, 
  J. Phys. Soc. Jpn. 22, 219 (1967).
\bibitem{degennes2} P. G. de Gennes, 
  in {\em Polymer Liquid Crystals}, 
  A. Ciferri, W. R. Krigbaum, R. Meyer eds. (Academic Press, New York, 1982).
\bibitem{warner} M. Warner, J. M. F. Gunn, A. B. Baumg\"artner,
  J. Phys. A {\bf 18}, 3007 (1985).
\bibitem{khodolenko} A. L. Khodolenko, T. A. Vilgis,
  Phys. Rev. E {\bf 52}, 3973 (1995).
\bibitem{oseen} C. Oseen, 
  Trans. Faraday Soc. {\bf 29}, 883(1933).
\bibitem{zocher} H. Z\"ocher,
  Trans. Faraday Soc. {\bf 29}, 945(1933).
\bibitem{frank} F. C. Frank, Discuss. Faraday Soc. {\bf 25}, 19 (1958).
\bibitem{flory1} P. J. Flory, 
  J. Chem. Phys. {\bf 9}, 660 (1941).
\bibitem{huggins} H. L. Huggins,
  J. Chem. Phys. {\bf 9}, 440 (1941).
\bibitem{flory2} P. J. Flory,
  {\em Principles of Polymer Chemistry}
  (Cornell University Press, Ithaca, New York, 1971).
\bibitem{footnote1}
  The length unit is $[x] = \sqrt{v_s/k_B T \: (K_{33}+K_{11})/2}$,
  and the energy unit is $[E] = k_B T/v_s \: [x^3]$.
\bibitem{footnote2}
  In the case of homeotropic substrate anchoring, $W < 0$, 
  a sharp tilting transition in the brush is still possible.
\end{thebibliography}
\end{document}